# Time course of oxidative damage in different brain regions following transient cerebral ischemia in gerbils

Authors: Eduardo Candelario-Jalil *; Noël H. Mhadu; Said M. Al-Dalain; Gregorio Martínez and Olga Sonia León

Affiliation: Department of Pharmacology, University of Havana (CIEB-IFAL), Apartado Postal 6079, Havana City 10600, Cuba.

*Author to whom all correspondence should be addressed:

**Eduardo Candelario-Jalil, M.Sc.**

**Department of Pharmacology**

**University of Havana (CIEB-IFAL)**

**Apartado Postal 6079**

**Havana City 10600**

**CUBA**

Phone: +53-7-219-536

Fax: +53-7-336-811

E-mail: candelariojalil@yahoo.com

**ABSTRACT**

The time course of oxidative damage in different brain regions was investigated in the gerbil model of transient cerebral ischemia. Animals were subjected to both common carotid arteries occlusion for 5 min. After the end of ischemia and at different reperfusion times (2, 6, 12, 24, 48, 72, 96 hr and 7 days), markers of lipid peroxidation, reduced and oxidized glutathione levels, glutathione peroxidase, glutathione reductase, manganese-dependent superoxide dismutase (MnSOD) and copper/zinc containing SOD (Cu/ZnSOD) activities were measured in hippocampus, cortex and striatum. Oxidative damage in hippocampus was maximal at late stages after ischemia (48 to 96 h) coincident with a significant impairment in glutathione homeostasis. MnSOD increased in hippocampus at 24, 48 and 72 h after ischemia, coincident with the marked reduction in the activity of glutathione-related enzymes. The late disturbance in oxidant-antioxidant balance corresponds with the time course of delayed neuronal loss in the hippocampal CA1 sector. Cerebral cortex showed early changes in oxidative damage with no significant impairment in antioxidant capacity. Striatal lipid peroxidation significantly increased as early as 2 h after ischemia and persisted until 48 h with respect to the sham-operated group. These results contribute significant information on the timing and factors that influence free radical formation following ischemic brain injury, an essential step in determining effective antioxidant intervention.

**Acknowledgements:**

The authors are greatly indebted to Dr. Joe E. Springer (Department of Anatomy and Neurobiology, University of Kentucky, Lexington, KY, USA) and Dr. Matthew B. Grisham (University of Louisiana Medical Center, Shreveport, LA, USA) for critically reading our manuscript. These studies were supported by Randox Laboratories (Antrim, UK).

## 1. INTRODUCTION

Cerebral ischemia results in a cascade of events leading to a number of important cellular changes. These include rapid decreases in ATP, calcium release from intracellular stores, loss of ion homeostasis, excitotoxicity, activation of enzymes (phospholipases, proteases, protein kinases, nitric oxide synthases, endonucleases), arachidonic acid release and metabolism, mitochondrial dysfunction, acidosis and edema (Macdonald and Stoodley, 1998; Lee et al., 1999). Many of these changes are associated with increased reactive oxygen species (ROS) production that can occur both during ischemia and at reperfusion. The role of oxidative stress becomes much greater in the case where cerebral blood flow is restored, because reflow to previous ischemic brain results in an increase in oxygen level, and consequently causes severe oxidative injury to the tissue by massive production of ROS (Chan, 1996). However, reperfusion is necessary to salvage the compromised ischemic tissue.

Oxidative stress is one of the most important factors that exacerbate brain damage by reperfusion. A large body of experimental research clearly shows that ischemia-reperfusion injury involves oxidatively damaging events (Cao et al., 1988; Kitagawa et al., 1990; Facchinetti et al., 1998). The brain is particularly vulnerable to oxidative injury because of its high rate of oxidative metabolic activity, intense production of reactive oxygen metabolites, high content of polyunsaturated fatty acids, relatively low antioxidant capacity, low repair mechanism activity and non-replicating nature of its neuronal cells (Evans, 1993).

Forebrain ischemia-reperfusion in gerbil is a model for human cerebral ischemia resulting from transient cardiac arrest. Certain brain regions, such as the striatum, neocortex and particularly the hippocampus, are more susceptible to ischemic damage (Kindy et al., 1992). In the hippocampus, the cornu Ammonis 1 (CA1) pyramidal neurons undergo selective delayed death several days after the injury (Kirino, 1982; Nitatori et al., 1995; Rao et al., 2000). Several lines of evidence indicate that oxidative stress contributes to delayed neuronal death after global cerebral ischemia (Kitagawa et al., 1990; Hall et al., 1993; Oostveen et al., 1998), suggesting that ROS formation may cooperate in a series of molecular events that link ischemic injury to neuronal cell death.

Thus, elucidation of the extent and the role of oxidative stress in the brain after ischemia-reperfusion are of great importance. A better understanding of the timing and factors that influence ROS formation is required for effective antioxidant intervention and for enlarging our knowledge of the pathophysiological mechanisms of cerebral ischemia. The temporal profile of histopathological changes in the gerbil brain following global ischemia has been extensively characterized, showing no neuronal loss up to 2-3 days of reperfusion but an extensive delayed neuronal loss at 5-7 days of reperfusion in the hippocampal CA1 region (Kirino, 1982; Rao et al., 2000; Martínez et al., 2001; Candelario-Jalil et al., unpublished data). Although ROS have been postulated to play an important role in the progression of reperfusion injury, the time course of oxidative damage following transient forebrain ischemia has been poorly characterized. In the present study, we have examined the time course of oxidative injury in different brain regions following transient global cerebral ischemia in gerbils. Further, the antioxidant capacity in each brain area was studied at different sampling times after the ischemic insult in view that oxidative stress may result not only from an increase in free radical production but also from a decrease in cellular antioxidant mechanisms. To our knowledge, the time course of the activity of glutathione-related enzymes as well as the content of both GSH and GSSG following transient forebrain ischemia in gerbils had not been previously characterized.

## 2. MATERIALS AND METHODS

### 2.1. Transient forebrain ischemia

Studies were performed in accordance with the Declaration of Helsinki and with the Guide for the Care and Use of Laboratory Animals as adopted and promulgated by National Institutes of Health (Bethesda, MD, USA). The experimental protocol was approved by our institutional animal care and use committee. Male Mongolian gerbils (*Meriones unguiculatus*; Hoe Gerk jirds strain) weighing 60-75 g at the time of surgery were used in this study. These animals were housed five per cage, exposed to a 12-h light/dark cycle, and had free access to food and water throughout the study period. The gerbils were anesthetized with chloral hydrate (300 mg/kg, i.p.). In the supine position, a midline ventral incision was made in the neck. Both common carotid arteries were exposed, separated carefully from the vagus nerve, and occluded for 5 minutes with microaneurysmal clips, which consistently resulted in delayed neuronal death in the CA1 region of the hippocampus (Kirino, 1982; Martínez et al., 2001). Blood flow during the occlusion and reperfusion after removal of the clips was visually confirmed and the incision was closed. The rectal temperature was monitored and maintained at 37 ± 0.5°C using an incandescent lamp and the animals were allowed to recover on an electrical heated blanket. In sham-operated group (n=5), the arteries were freed from connective tissue but were not occluded.

At the end of ischemia (no reflow; n=5) and at different reperfusion times (2, 6, 12, 24, 48, 72, 96 hr and 7 days), animals (n=6 per each time point) were deeply anesthetized with diethyl ether and perfused intracardially with ice-cold saline to flush all blood components from the vasculature. Brains were quickly removed, kept in ice-cold saline and immediately dissected on a cold plate using the atlas of Paxinos and Watson (1998) as a reference, exactly as in our previous experiments (Candelario-Jalil et al., 2000; Martínez et al., 2001). Three different regions (hippocampus, cortex and striatum) were removed, weighed and homogenized in ice-cold 20 mM Tris-HCl buffer (pH 7.4) and centrifuged for 10 min at 12,000 x *g*. The supernatant was collected, frozen at -20°C and employed for biochemical analyses.

It is possible that isolation and processing of the brain tissue may result in production of free radicals and subsequent oxidative injury that is not specific to ischemic insult. However, the use of ice-cold homogenization buffer and the fact that all samples were handled and processed in an identical fashion reduces the potential for generating non-specific free radical species. In addition, all animals were perfused transcardially with ice-cold saline in order to eliminate the excess of iron (bound to hemoglobin) that may artificially increase free radical formation via Fenton reaction.

*2.2. Lipid peroxidation assays*
Lipid peroxidation (LP) was assessed by measuring the concentration of malondialdehyde (MDA) and 4-hydroxyalkenals (4-HDA) and by determining the levels of lipid hydroperoxides in brain samples. Concentrations of MDA and 4-HDA were analyzed using the LPO-586 kit obtained from Calbiochem (La Jolla, CA). In the assay, the production of a stable chromophore after 40 min of incubation at 45°C was measured at a wavelength of 586 nm. For standards, freshly prepared solutions of malondialdehyde bis [dimethyl acetal] (Sigma) and 4-hydroxynonenal diethylacetal (Cayman Chemical) were employed and assayed under identical conditions. Concentrations of MDA and 4-HDA in brain samples were calculated using the corresponding standard curve and values were expressed as nmol MDA+4-HDA per mg protein. This procedure has been used widely for the measurement of products of LP in brain homogenates (Melchiorri et al., 1996; Chabrier et al., 1999; Candelario-Jalil et al., 2000). Lipid hydroperoxides were measured with ferrous oxidation-xylenol orange assay (Gay et al., 1999) as reported for brain homogenates (Song et al., 1999). Hydrogen peroxide was used as reference standard (R=0.996). Lipid hydroperoxides levels were expressed as nmol hydroperoxides per mg protein.

*2.3. Glutathione determination*

Reduced and oxidized glutathione (GSH and GSSG, respectively) were measured enzymatically in 5-sulphosalycilic acid-deproteinized samples by using a modification (Anderson, 1985) of the procedure of Tietze (1969) as described for brain samples (Floreani et al., 1997). The method is based on the determination of a chromophoric product, 2-nitro-5-thiobenzoic acid, resulting from the reaction of 5,5'-dithiobis(2-nitrobenzoic acid) with GSH. In this reaction, GSH is oxidized to GSSG, which is then reconverted to GSH in the presence of glutathione reductase (type III, from *Saccharomyces cerevisiae*, Randox Laboratories, Antrim, UK) and NADPH. The rate of 2-nitro-5-thiobenzoic acid formation, which is proportional to the sum of GSH and GSSG present is followed at 412 nm. Samples were assayed rapidly to minimize GSH oxidation. Specificity of this method for glutathione quantification is ensured by highly specific glutathione reductase. GSH present in the samples was calculated as the difference between total glutathione and GSSG levels, taking into account the fact that one molecule of GSSG gives rise to two molecules of GSH upon reaction with glutathione reductase. A standard curve with known amounts of GSH was established and employed for estimating glutathione content.

*2.4. Antioxidant enzymes assays*

Glutathione peroxidase (GPx) activity was assayed using a commercial kit obtained from Randox Laboratories (Antrim, UK), which is based on the procedure described by Flohé and Gunzler (1984) using cumene hydroperoxide as substrate. The reaction was followed for 3 min at 340 nm in a Pharmacia LKB Ultraspec Plus spectrophotometer. The contribution of spontaneous NADPH oxidation was always subtracted from the overall reaction rate. GPx activity was expressed as nmol NADPH oxidized per minute per mg protein.

Glutathione reductase (GR) activity was determined according to Carlberg and Mannervik (1985). The oxidation of NADPH was followed for 3 min at 340 nm and the activity of GR was calculated using a molar extinction coefficient of 6.3 $mM^{-1}cm^{-1}$. Non-enzymatic NADPH oxidation was subtracted from the overall rate. GR activity was expressed as nmol NADPH oxidized per minute on the basis of total protein content.

Superoxide dismutase (SOD) was measured using pyrogallol as substrate (Shukla et al., 1987). This method follows the superoxide-driven auto-oxidation of pyrogallol at pH 8.2 in the presence of EDTA. The assay mixture contained 1 mM EDTA in 50 mM Tris-HCl buffer (pH 8.2) with or without the sample. The reaction was started by the addition of pyrogallol (final concentration 0.124 mM) and the oxidation of pyrogallol was followed for 1 min at 420 nm. The percent inhibition of the auto-oxidation of pyrogallol by SOD present in the tissue sample was determined, and standard curves using known amounts of purified SOD (Sigma) under identical conditions were established. One unit (U) of SOD activity was defined as the amount that reduced the absorbance change by 50%, and results were normalized on the basis of total protein content (U/mg protein). Copper/zinc superoxide dismutase (Cu/ZnSOD) was differentiated from manganese superoxide dismutase (MnSOD) by addition of 2 mM sodium cyanide to inhibit the activity of Cu/ZnSOD from total SOD activity. Cu/Zn SOD activity was calculated as the difference between total SOD and MnSOD activity as in a previous report (McIntosh et al., 1998).

*2.5. Protein assay*

Total protein concentrations were determined using the method described by Bradford (1976) and analytical grade bovine serum albumin was used to establish a standard curve. Chemicals and reagents were purchased from Sigma Chemical Co. (Saint Louis, MO, USA).

*2.6. Statistical analysis*

Data are expressed as mean ± standard deviation. Statistical analysis was performed with one-way ANOVA followed by a Student-Newman-Keuls post-hoc test. The value of *P* less than 0.05 was considered to be statistically significant.

**3. RESULTS**

*3.1. Effects of cerebral ischemia on lipid peroxidation markers*

In hippocampus, bilateral carotid occlusion for 5 min in gerbils resulted in marked increase in lipid peroxidation as shown in Figure 1. Malondialdehyde (MDA) and 4-hydroxyalkenals (4-HDA) levels significantly (p<0.05) increased at 6 h of reperfusion and remained high until 96 h when compared to sham-operated animals. In a similar way, lipid hydroperoxides increased as early as 2 h of recirculation, remained increased until 12 h, trended downward by 24 h, but significantly increased at 48, 72 and 96 h of reflow.

Cortical MDA and 4-HDA concentrations were significantly ($p<0.05$) elevated after 12 and 24 h of reperfusion as compared to non-ischemic control group. Lipid hydroperoxides levels increased with statistical significance ($p<0.05$) after 6 h of recirculation and remained high until 24 h, similarly to MDA and 4-HDA concentrations. In cerebral cortex, both lipid peroxidation markers subsided at 48 h and remained at basal levels until the last sampling time (7 d). In striatum, both lipid hydroperoxides and MDA + 4-HDA concentrations significantly increased at 2 and 6 h of reperfusion respectively, and were kept elevated until 48 h with respect to the sham group (Fig. 1).

On the other hand, no significant increase in lipid peroxidation was found in any brain region at 7 days of reperfusion when compared to control animals. In a like manner, brief ischemia (5 min) without reperfusion failed to produce increases in lipid peroxidation indices (Fig. 1).

*3.2. Time course of ischemia-induced modifications of glutathione levels in the gerbil brain*

Hippocampal GSH content significantly decreased ($p<0.05$) in the early recirculation periods (2 and 6 h) after ischemia, returned to that of control group at 12 and 24 h of reflow and reached maximal reductions at 48 and 72 h with respect to sham-operated gerbils (Table 1). GSSG levels were elevated with respect to control values at 2, 6, 24, 48 and 72 h of reperfusion as shown in Table 1. Cortical GSH only showed a significant decrease at 6 h after 5 min of transient ischemia, but normalized to control levels thereafter. By contrast, GSSG content increased with statistical significance ($p<0.05$) after 2, 6, 12 and 24 h of reflow with respect to the sham-operated group (Table 1). In corpus striatum, GSH levels remained constant at all times tested. However, striatal GSSG significantly increased between 2 and 48 h of reflow in comparison with the control group (Table 1). There were no changes in GSH or GSSG in the ischemic only gerbils as compared to the sham-operated animals in any brain region.

*3.3. Effects of ischemia on antioxidant enzymes*

Figure 2 shows that hippocampal glutathione peroxidase activity significantly decreased by 22 and 18% versus sham-operated group after 48 and 72 h of recirculation respectively. In hippocampus, glutathione reductase activity was significantly reduced by 46, 42 and 31% when compared with the sham group at 48, 72 and 96 h of reperfusion respectively. In contrast, no statistically significant differences in glutathione-related enzyme activities were observed at any time point in the other brain regions (data not shown). Finally, the effects of ischemia and different reperfusion times on both MnSOD and Cu/ZnSOD activities in hippocampus, cortex and striatum are presented in Table 2. Hippocampal MnSOD activity significantly increased at 24, 48 and 72 h of reperfusion (57, 46 and 51% vs. control, respectively). MnSOD activity did not change in cerebral cortex or striatum at any time point examined. On the other hand, striatal Cu/ZnSOD activity was significantly reduced at 12 h of reperfusion following ischemia. A similar trend was observed at 24 h in this brain area although this reduction did not reach statistical significance (Table 2). However, no significant changes in Cu/ZnSOD activity were seen at any reperfusion time in either hippocampus or cortex. Moreover, a brief period of ischemia without reflow resulted in no significant changes in glutathione-related enzymes or SODs activities.

## 4. DISCUSSION

There is increasing evidence that the brain damage produced by cerebral ischemia develops over a period longer than previously believed. A critical role of oxidative stress has been implicated in ischemic brain damage. The findings of this study show that certain neuronal populations are highly susceptible to oxidative damage induced by a brief global cerebral ischemia episode, which is very similar to the pattern of neuronal loss assessed histopathologically (Kirino, 1982; Nitatori et al., 1995). The time course of oxidative injury is not the same in the different brain regions examined. In hippocampus, it was observed a late maximal increase in markers of oxidative damage coincident with a significant impairment of glutathione homeostasis. On the other hand, cerebral cortex showed early changes in oxidative damage with no significant impairment in antioxidant capacity. Unlike cortex, striatal oxidative damage persisted until 48 h of reflow showing reduction in Cu/ZnSOD activity but no late alterations in glutathione homeostasis.

*4.1. Effects of ischemia on markers of lipid peroxidation in hippocampus*
Strong experimental evidences suggest that lipid peroxidation plays a widespread role in neuronal cell death. Lipid peroxidation results in loss of membrane integrity, impairment of the function of membrane-transport proteins and ion channels, disruption of cellular ion homeostasis and concomitantly increases neuronal vulnerability to excitotoxicity (Mattson, 1998; Springer et al., 1997).

In hippocampus, our results have shown a sustained elevation of markers of lipid peroxidation following transient forebrain ischemia, which was maximal at 48, 72 and 96 h of recirculation. Our findings are consistent with other studies, which have found that the increased level of lipid peroxidation persists for several days after brief forebrain ischemia in the gerbil hippocampus (Floyd and Carney, 1991; Haba et al., 1991; Caldwell et al., 1995). We have employed two distinct methods for assessing lipid peroxidation. Both MDA + 4-HDA and lipid hydroperoxides levels were similarly increased at the same time points. Agreement of both assays for assessing lipid peroxidation provides assurance of the involvement of lipid peroxidative mechanisms in global cerebral ischemia in our experimental conditions. Other studies have shown a significant delayed increase in MDA-related immunostaining at 48 and 72 h in the gerbil brain subjected to a 5-min episode of near-complete global ischemia (Hall et al., 1997; Oostveen et al., 1998). In a more recent report, 4-hydroxy-2-nonenal immunoreactivity in CA1 pyramidal neurons increased markedly from 8 h to seven days after transient global ischemia in gerbils (Urabe et al., 2000). It is interesting to note that the time course of increased lipid peroxidation was simultaneous with that of post-ischemic CA1 neuronal degeneration (Hall et al., 1997). Additionally, a recent study has shown a significant delayed increase in superoxide radical generation in hippocampus on the third and fifth days of reperfusion following 4 min of transient ischemia in gerbils, which correlated with histopathological changes in the CA1 hippocampal sector (Yamaguchi et al., 1998).

Then, it seems reasonable to assume that the delayed occurrence of oxidative injury may correlate well with delayed neuronal loss of hippocampal CA1 pyramidal neurons. Nevertheless, the mechanisms underlying the increase in oxidative damage at late stages after the initial ischemic insult are not completely understood. It was recently found that glial (GLT-1 and GLAST) and neuronal (EAAC1) high-affinity glutamate reuptake mechanisms are downregulated at late stages after ischemia, which precedes delayed neuronal death in gerbil hippocampus (Rao et al., 2000). Their dysfunction leads to neuronal damage by allowing glutamate to remain in the synaptic cleft for a longer duration and so excitotoxicity-induced oxidative injury is likely to occur in the ischemic brain. Other mechanisms which might account for the late increase in oxidative damage are probably the delayed induction of ROS-generating enzymes like cyclooxygenase-2 (Ohtsuki et al., 1996) and nitric oxide synthase (Yrjänheikki et al., 1998).

*4.2. Effects of ischemia on hippocampal antioxidant defenses*

In an attempt to further elucidate the factors that might influence ROS formation following cerebral ischemia, we have measured the antioxidant capacity in each brain area at different reflow times following ischemia. Our results have shown a significant reduction of both glutathione peroxidase (GPx) and glutathione reductase (GR) activity in the hippocampus at late periods of reperfusion, coincident with maximal reduction in GSH and marked increase in GSSG (Table 1). The most robust and significant alteration in the antioxidant defense is a decrease in GSH content (Schulz, 2000). The persistent and profound decrease in hippocampal GPx and GR activities as well as the marked decrease in GSH levels following ischemia (Fig. 2 and Table 1) is indicative of the lowered antioxidant capacity of this brain region being possibly related to their greater vulnerability toward ischemia. Our results suggest that the delayed impairment of hippocampal glutathione homeostasis might be involved in the late increase in oxidative damage, which is likely related to delayed neuronal loss in this brain area. Oxygen radicals have been reported to inactivate GR (Huang and Philbert, 1996) and GPx (Pigeolet and Remacle, 1991). In addition, depletion of brain GSH results in a decrease of GR activity, an enzyme susceptible to oxidative injury (Barker et al., 1996). It is important to note that the marked reduction in GR and GPx activity occurred when GSH depletion and LP were maximal (Figs. 1 and 2; Table 1). Additionally, GSSG concentrations in hippocampus showed their maximal increase at 48 and 72 h, coincident with the maximal reduction in GR activity. The timing of the observed ischemia-induced maximal decrease in GSH (48-72 h after ischemia) indicates that the latter may represent an index of neuronal damage prior to death. Loss of glutathione and oxidative damage have been suggested to constitute early, possibly signaling events in apoptotic cell death (Sato et al., 1995; Chandra et al., 2000), although GSH depletion alone may not trigger apoptosis (Wüllner et al., 1999). However, low cellular GSH levels could allow oxidative stress to occur which might favor the onset of apoptosis.

On the other hand, hippocampal manganese superoxide dismutase (MnSOD) activity showed a sustained elevation between 24 and 72 h of reperfusion, most likely indicating a protective response to heightened oxidative stress after the initial ischemic insult. SOD is essential for removal of superoxide radicals ($O_2 \cdot -$) from cells. The removal of $O_2 \cdot -$ by SOD prevents the production of the hydroxyl radical ($\cdot OH$); although paradoxically, the $H_2O_2$ it produces can interact with metal ions ($Fe^{2+}$, $Cu^+$) to produce $\cdot OH$ via Fenton reaction. Hence, SOD is beneficial only in the presence of sufficient $H_2O_2$-detoxifying enzymes, such as catalase and GPx. In view that catalase activity is very low in brain, GPx is the main hydroperoxide-detoxifying enzyme in the central nervous system (Halliwell and Gutteridge, 1985). It is probably that delayed increase in MnSOD activity (Table 2) in the hippocampus is not beneficial in view that GPx activity is not concomitantly increased. On the contrary, GPx is significantly reduced at the late stages of reflow, which possibly accounts for the marked increase in lipid hydroperoxides observed at these reperfusion times (Fig. 1). It is possible that all these factors contribute to the cascade of events that ultimately mediate neuronal death after transient forebrain ischemia.

*4.3. Oxidative stress in cerebral cortex and striatum following forebrain ischemia*

Unlike hippocampus, cerebral cortex showed early changes in markers of oxidative damage but these indices subsided after 24 h of reperfusion. Ischemia was unable to significantly alter enzymatic antioxidant defenses in cerebral cortex. The mechanisms responsible for this highly region-specific pattern of oxidative damage induced by ischemia are far from being well understood. On the other hand, in the striatum, the ischemia-induced increases in markers of oxidative injury persisted over longer periods of time after ischemia. This susceptibility might be associated with dopamine oxidation (either enzymatic or non-enzymatic) that increases the production of reactive oxygen species. This notion is based in previous reports which have demonstrated that prior depletion of dopamine is neuroprotective in cerebral ischemia suggesting that ischemia-induced dopamine release may contribute to increased oxidative stress in the striatum (Globus et al., 1987; Ren et al., 1997).

*4.4. Concluding remarks*

In summary, the results of this study show that transient ischemia-induced oxidative injury evolves temporally and spatially, and provide further evidence that oxidative stress may be involved in delayed neuronal death following ischemia in gerbils. In addition, our present findings suggest that the disturbance in oxidant-antioxidant balance at late stages after the ischemic insult might play a part in rendering brain tissue more vulnerable to free radical-induced injuries. These results might have important implications for the antioxidant therapy in transient global cerebral ischemia due to the maximal appearance of reactive oxygen species in the late stages, which suggest that postischemic administration of antioxidants would be probably valuable in preventing hippocampal neuronal loss.

**Table 1.** Reduced (GSH) and oxidized glutathione (GSSG) levels in different brain regions at different time intervals after 5 min of transient global cerebral ischemia in gerbils.

| | Hippocampus | | Cortex | | Striatum | |
|---|---|---|---|---|---|---|
| Time, h | GSH (μg/g tissue) | GSSG (ng/g tissue) | GSH (μg/g tissue) | GSSG (ng/g tissue) | GSH (μg/g tissue) | GSSG (ng/g tissue) |
| Sham | 1.41 ± 0.13 [a] | 1.45 ± 0.46 [a] | 1.42 ± 0.17 [a] | 1.53 ± 0.25 [a] | 1.33 ± 0.12 [a] | 1.44 ± 0.43 [a] |
| Ischemia | 1.50 ± 0.28 [a] | 1.67 ± 0.51 [a] | 1.45 ± 0.11 [a] | 1.86 ± 0.99 [a] | 1.28 ± 0.2 [a] | 1.56 ± 0.62 [a] |
| 2 | 1.00 ± 0.13 [b,c] | 6.35 ± 1.56 [d] | 1.13 ± 0.19 [a,b] | 3.94 ± 0.88 [c] | 1.11 ± 0.18 [a] | 2.91 ± 0.67 [d] |
| 6 | 1.09 ± 0.10 [c] | 7.84 ± 1.90 [d] | 1.01 ± 0.14 [b] | 7.64 ± 0.97 [b] | 0.87 ± 0.31 [a] | 6.65 ± 0.87 [b] |
| 12 | 1.45 ± 0.18 [a] | 2.61 ± 1.26 [a] | 1.1 ± 0.11 [a,b] | 4.58 ± 1.19 [c] | 1.04 ± 0.23 [a] | 5.7 ± 0.6 [c] |
| 24 | 1.20 ± 0.19 [a,c] | 7.23 ± 2.18 [d] | 1.20 ± 0.1 [a,b] | 4.16 ± 1.46 [c] | 1.09 ± 0.47 [a] | 5.74 ± 0.84 [c] |
| 48 | 0.80 ± 0.17 [b] | 16.5 ± 2.60 [b] | 1.33 ± 0.23 [a] | 1.78 ± 0.42 [a] | 1.21 ± 0.13 [a] | 3.4 ± 0.99 [d] |
| 72 | 1.10 ± 0.11 [c] | 12.97 ± 2.4 [c] | 1.40 ± 0.13 [a] | 2.03 ± 0.99 [a] | 1.31 ± 0.17 [a] | 1.35 ± 0.32 [a] |
| 96 | 1.29 ± 0.32 [a,c] | 4.26 ± 1.26 [a] | 1.47 ± 0.10 [a] | 1.84 ± 0.54 [a] | 1.33 ± 0.16 [a] | 1.66 ± 0.41 [a] |
| 7 d | 1.50 ± 0.14 [a] | 1.50 ± 0.39 [a] | 1.35 ± 0.37 [a] | 1.69 ± 0.6 [a] | 1.29 ± 0.17 [a] | 1.49 ± 0.63 [a] |

Data are mean ± SD. Means having different superscript letters indicate significant difference ($P<0.05$) between groups.

**Table 2.** Effects of 5 min of transient global cerebral ischemia on manganese-dependent and copper/zinc-dependent superoxide dismutase activity (MnSOD and Cu/ZnSOD, respectively) in different gerbil brain regions.

|  | **Hippocampus** | | **Cortex** | | **Striatum** | |
|---|---|---|---|---|---|---|
|  | **U/mg protein** | | **U/mg protein** | | **U/mg protein** | |
| **Time (h)** | **MnSOD** | **Cu/ZnSOD** | **MnSOD** | **Cu/ZnSOD** | **MnSOD** | **Cu/ZnSOD** |
| **Sham** | 3.7 ± 0.9 | 6.5 ± 0.8 | 2.6 ± 0.7 | 4.8 ± 0.8 | 3.01 ± 0.5 | 5.9 ± 0.6 |
| **Ischemia** | 3.6 ± 0.8 | 6.3 ± 1.1 | 2.5 ± 0.5 | 4.9 ± 0.9 | 2.9 ± 0.5 | 6.2 ± 1.0 |
| **2** | 3.5 ± 0.5 | 6.6 ± 0.7 | 2.4 ± 0.3 | 4.9 ± 0.5 | 2.7 ± 0.6 | 5.7 ± 1.0 |
| **6** | 3.3 ± 1.5 | 6.6 ± 1.3 | 2.8 ± 0.6 | 5.0 ± 0.9 | 3.3 ± 0.9 | 5.3 ± 0.8 |
| **12** | 3.6 ± 0.9 | 6.2 ± 1.0 | 2.6 ± 0.8 | 5.1 ± 1.0 | 3.1 ± 0.4 | 4.2 ± 0.4 * |
| **24** | 5.8 ± 0.7 * | 6.6 ± 1.4 | 2.5 ± 0.4 | 5.3 ± 0.8 | 3.6 ± 0.7 | 4.5 ± 0.8 |
| **48** | 5.4 ± 0.6 * | 6.4 ± 0.9 | 2.6 ± 0.4 | 5.0 ± 0.7 | 3.2 ± 0.5 | 5.6 ± 1.0 |
| **72** | 5.6 ± 1.1 * | 6.6 ± 0.8 | 2.6 ± 0.28 | 5.1 ± 0.6 | 3.4 ± 0.7 | 6.1 ± 1.2 |
| **96** | 3.9 ± 0.7 | 7.1 ± 0.7 | 2.9 ± 0.9 | 4.9 ± 1.1 | 2.9 ± 0.6 | 6.3 ± 1.3 |
| **7 d** | 3.8 ± 0.8 | 7.0 ± 1.6 | 3.0 ± 1.2 | 5.1 ± 1.8 | 3.3 ± 0.9 | 6.1 ± 0.9 |

Data are mean ± SD.

* $P<0.05$ with respect to sham-operated animals.

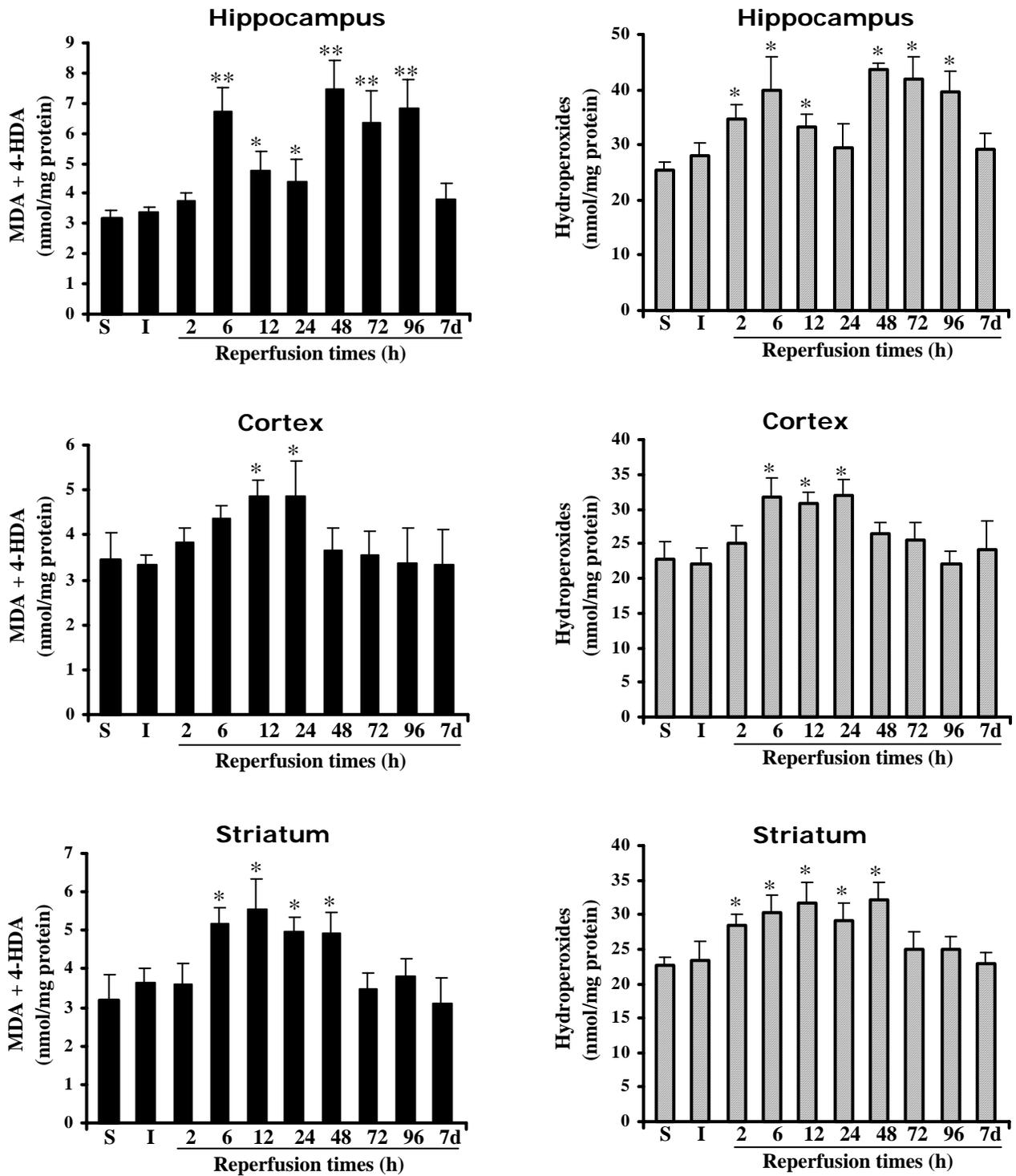

**Fig. 1.** Lipid peroxidation, as assessed by malondialdehyde (MDA), 4-hydroxyalkenals (4-HDA) and lipid hydroperoxides levels in different brain areas at different time intervals after 5 min of transient forebrain ischemia in gerbils. Data are mean ± SD. *$P<0.05$ and **$P<0.01$ with respect to sham-operated animals. S: Sham group; I: Ischemia without reperfusion.

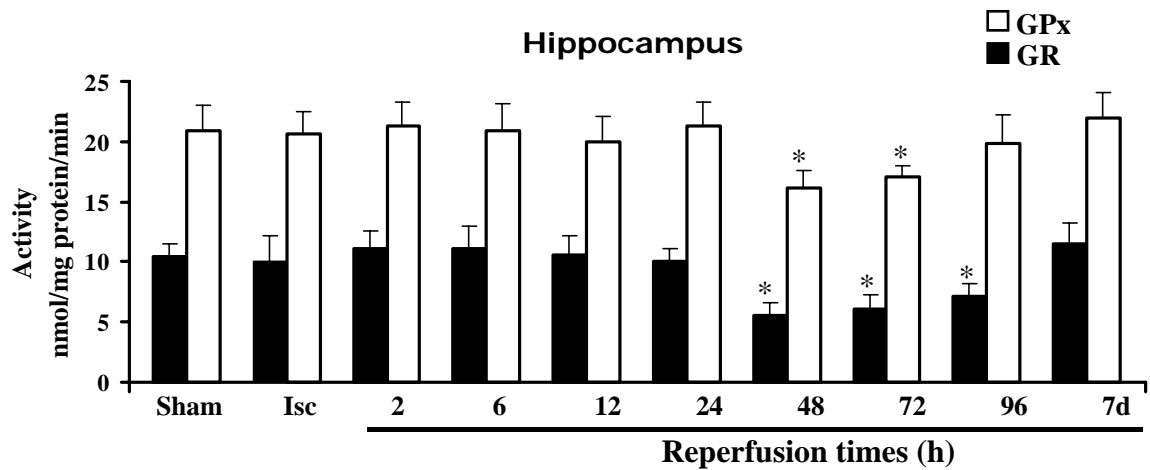

**Fig. 2.** Effects of ischemia on hippocampal glutathione peroxidase (GPx) and glutathione reductase (GR) activities at different reperfusion times following 5 min of global cerebral ischemia in the gerbil. Data are mean ± SD. *$P<0.05$ with respect to sham-operated animals. S: Sham group; I: Ischemia without reperfusion.